\documentclass[graybox,envcountchap,sectrefs]{svmult}

\usepackage{mathptmx}
\usepackage{helvet}
\usepackage{courier}
\usepackage{braket} 
\usepackage{algorithmicx}
\usepackage{algpseudocode} 
\usepackage{amsfonts}

\usepackage{type1cm}         
\usepackage{exercise}
\usepackage{makeidx}         
\usepackage{graphicx}        
\usepackage{multicol}        
\usepackage[bottom]{footmisc}
\usepackage[usenames,dvipsnames,x11names]{xcolor}
 \usepackage{listings}
 \usepackage{epic}
 \usepackage{eepic}
 \usepackage{a4wide}
 \usepackage{color}
 \usepackage{amsmath}
 \usepackage{amssymb}
 \usepackage[T1]{fontenc}
 \usepackage{cite} 
 \usepackage{shadow}
 \usepackage{hyperref}
 \usepackage{bezier}
 \usepackage{pstricks}
\usepackage{textcomp,type1ec,pdfpages}
\usepackage{bera}

\makeindex             

\usepackage{hyperref}

\begin{document}



\setcounter{chapter}{5}

\title{Lattice methods and the nuclear few- and many-body problem}
\author{Dean Lee}
\institute{Dean Lee \at  Department of Physics, 
North Carolina State University, Raleigh, NC 27695,  USA, \email{dean\_lee@ncsu.edu}}
\maketitle

\abstract{We begin with a brief overview of lattice calculations using chiral effective field theory and some recent applications.  We then describe several methods for computing scattering on the lattice.  After that we focus on the main goal, explaining the theory and algorithms
relevant to lattice simulations of nuclear
few- and many-body systems.  We discuss the exact equivalence
of four different lattice formalisms, the Grassmann path integral, transfer matrix operator, Grassmann path
integral with auxiliary fields, and transfer matrix operator with auxiliary
fields.  Along with our analysis we include several coding examples and a number of exercises for the calculations of few- and many-body systems at leading order in chiral effective field theory.}

\section{Introduction}
This chapter builds upon the general overview of lattice methods for effective field theory of the previous chapter. We discuss the theory and algorithms used in lattice simulations of nuclear
few and many body systems.  We show the exact equivalence of the Grassmann path integral, transfer matrix operator, Grassmann path
integral with auxiliary fields, and transfer matrix operator with auxiliary
fields.  Along with our analysis we include several coding examples and a number of exercises for the calculations of few- and many-body systems at leading order in chiral effective field theory.

Effective field theory (EFT) provides a theoretical framework for organizing low-energy
interactions in powers of particle
momenta. \ Chiral
effective field theory   applies this framework to the low-energy interactions of protons and neutrons while explicitly including the interactions of pions \cite{Weinberg:1990rz,Weinberg:1991um,Ordonez:1992xp,Ordonez:1993tn,vanKolck:1994yi,Epelbaum:1998hg,Epelbaum:1998ka,Bedaque:2002mn,Epelbaum:2008ga}.
Pions are qualitatively different from other mesons since they become massless in the limit of massless quarks, thereby producing long-range exchange interactions. The low-energy expansion of chiral EFT is organized
in powers of $Q$, where $Q$ denotes the typical momentum of the nucleons as well as explicit factors of the pion mass. \ The most important interactions are called
leading order (LO) or $O(Q^{0})$. \ The next most important contributions
are 
next-to-leading order (NLO) or $O(Q^{2})$. \ The terms after this are
next-to-next-to-leading order (NNLO) or $O(Q^{3})$, and so on. 

Lattice EFT refers generally to lattice simulations based upon the framework of effective
field theory.  There are a few reviews in the literature which discuss current methods used
in lattice effective field theory \cite{Lee:2008fa,Drut:2012a} as well as the discussion in the previous chapter of this volume. Many different phenomena can be studied
in
lattice EFT using the same lattice action. \ In principle all systematic
errors are introduced up front when defining the low-energy effective theory,
as opposed to the particular computational scheme used to calculate
observables. \ 

Lattice EFT has been aided by efficient lattice methods developed for lattice QCD and condensed matter applications. \ The methods include
Markov Chain Monte Carlo techniques, auxiliary fields
\cite{Hubbard:1959ub,Stratonovich:1958}, pseudofermion methods
\cite{Weingarten:1980hx}, and non-local updating schemes such as the hybrid Monte
Carlo algorithm\cite{Scalettar:1986uy,Gottlieb:1987mq,Duane:1987de}. \ Lattice EFT
was
first used in studies of infinite nuclear matter \cite{Muller:1999cp} and
infinite neutron matter with and without explicit pions
\cite{Lee:2004si,Lee:2004qd,Lee:2005is,Lee:2005it}. \ The method has also been used
to
study light nuclei in pionless EFT \cite{Borasoy:2005yc} and chiral EFT at
leading order \cite{Borasoy:2006qn}. There have been further studies of neutron matter 
\cite{Borasoy:2007vi,Borasoy:2007vk,Wlazlowski:2014jna} and light nuclei \cite{Epelbaum:2009zs,Epelbaum:2009pd}, and there have been several applications to nuclear structure and nuclear clustering \cite{Epelbaum:2011md,Epelbaum:2012qn,Epelbaum:2012iu,Lahde:2013uqa,Epelbaum:2013paa,Elhatisari:2016owd} as well as recent work on nuclear scattering and reactions~\cite{Rupak:2013aue,Rupak:2014xza,Elhatisari:2015iga}.

\section{Recent Applications}

We review here several recent applications of lattice effective field theory to nuclear systems. In Ref.~\cite{Epelbaum:2013paa},
the first {\it ab initio} evidence is presented for a tetrahedral alpha-cluster structure
of the ground state of $^{16}$O. The first excited $0^+$ state of $^{16}$O is found to be a planar or square arrangement of alpha clusters.  The evidence for these geometric arrangements come from the strong overlap between nuclear states and initial state configurations with these alpha-cluster geometries.

In Table~\ref{oxygen1} we presented the energies of the low-lying even parity states of oxygen-16.  The columns labeled ``LO(2N)'' and ``NNLO(2N)'' show
the energies at each order using the two-nucleon force only. The column labeled
``+3N'' also includes the 3NF, which first appears 
at NNLO. The column ``+4N$_\mathrm{eff}$'' includes an ``effective''
4N force, and the column ``Exp'' gives
the empirical energies.  This ``effective'' 4N force was introduced in Ref.~\cite{Lahde:2013uqa} as a proxy  measure of unknown systematic errors responsible for overbinding in lattice chiral effective field theory calculations with increasing numbers of nucleons.  This tendency towards overbinding has also been noted in other nuclear structure calculations \cite{Ekstrom:2015rta,Hagen:2015yea}. 
\begin{table}[h]
\centering
\caption{Lattice results and experimental energies for the lowest even-parity
states of $^{16}$O in MeV. 
The errors include statistical
Monte Carlo errors and 
uncertainties due to the extrapolation to infinite Euclidean time. 
\label{oxygen1}}
\vspace{.5cm}
\begin{tabular}{c | r | r r r | r}
$J_n^p$ & \multicolumn{1}{c |}{LO (2N)} & \multicolumn{1}{c}{NNLO (2N)} 
& \multicolumn{1}{c}{+3N} & \multicolumn{1}{c |}{+4N$_\mathrm{eff}$} & \multicolumn{1}{c}{Exp}
  \\ \hline\hline
$0^+_1$ & $-147.3(5)$ & $-121.4(5)$ & $-138.8(5)$ & $-131.3(5)$ & $-127.62$
\\
$0^+_2$ & $-145(2)$ & $-116(2)$ & $-136(2)$ & $-123(2)$ & $-121.57$ \\
$2^+_1$ & $-145(2)$ & $-116(2)$ & $-136(2)$ & $-123(2)$ & $-120.70$
\end{tabular}
\end{table}

In order to understand the source of this overbinding, the problem was revisited again in Ref.~\cite{Elhatisari:2016owd}.  In that work numerical evidence from {\it ab initio} lattice simulations showed that the problem appears related to the fact that the nuclear forces reside near a quantum phase transition. Using lattice effective field theory, Monte Carlo simulations were performed for systems with up to twenty nucleons. For even and equal numbers of protons and neutrons, a first-order transition was found at zero temperature from a Bose-condensed gas of alpha particles to a nuclear liquid. Whether one has an alpha-particle gas or nuclear liquid is determined by the strength of the alpha-alpha interactions, and the alpha-alpha interactions depend on the strength and locality of the nucleon-nucleon interactions. This insight is useful in improving calculations of nuclear structure and important astrophysical reactions involving alpha capture on nuclei.  These findings also provide a tool to probe the structure of alpha cluster states such as the Hoyle state responsible for the production of carbon in red giant stars and point to a connection between nuclear states and the universal physics of bosons at large scattering length.

Processes such as the scattering of alpha particles,
the triple-alpha reaction, and
alpha capture play an important role in stellar nucleosynthesis.  In
particular, alpha capture on carbon determines the ratio of carbon to oxygen
during helium burning and impacts the following carbon, neon, oxygen,
and silicon burning stages.  In these
reactions the elastic scattering of alpha particles
and alpha-like nuclei (nuclei with even and equal numbers of protons
and neutrons) are important for understanding background and resonant
scattering contributions.  In Ref.~\cite{Elhatisari:2015iga}
the first {\it ab initio}
calculations of the scattering of two alpha particles were performed using a technique called the adiabatic projection method.  These
calculations represent a significant algorithmic improvement since the calculations presented in
 Ref.~\cite{Elhatisari:2015iga}
scale roughly quadratically with the number of nucleons and opens a gateway to scattering and reactions involving heavier nuclei. \\

\section{Scattering on the lattice}

At any given order in the chiral EFT expansion, there will be short-range interaction coefficients which depend on the chosen regularization of the large-momentum divergences.  On the lattice this regularization is provided by the lattice spacing, unless some additional regularization
is applied to the lattice interactions.
In order to set the values of the short-range two-nucleon interaction
coefficients, we make a comparison of nucleon-nucleon scattering on
the lattice with experimental scattering data.  The extension to three-nucleon
interaction coefficients is also required at NNLO, and that procedure
on the lattice has been discussed in Ref.~\cite{Epelbaum:2009zs}

As discussed in the previous chapter, L\"{u}scher \cite{Luscher:1985dn,Luscher:1986pf,Luscher:1991ux} has shown that the finite-volume energy levels for a two-body system in a periodic cubic box are related to the infinite-volume scattering matrix. \ While the method is very useful
at low momenta, it can become less accurate at higher momenta and higher orbital angular momenta. \ Also spin-orbit
coupling and partial-wave mixing are difficult to measure accurately using
L\"{u}scher's method due to scattering artifacts produced by the
cubic periodic boundary. \ An alternative approach has been developed to measure
phase shifts for particles on the lattice using a spherical wall boundary \cite{Borasoy:2007vy,Carlson:1984}.
\ 

In this approach, a hard spherical wall boundary is imposed on the relative separation between
the two particles.  This wall is placed at some chosen  radius $R_{\text{wall}}$, and it removes copies of the interactions produced by the periodic lattice.
\ Working in the center-of-mass frame, we solve the time-independent Schr{\"o}dinger equation as a function of the relative separation between the particles and compute spherical standing waves which vanish at $r=R_{\text{wall}}$.
At values of $r$ beyond the range of the interaction, the spherical standing
waves can be written as a superposition of products of spherical harmonics
and spherical Bessel functions,\begin{equation}
\left[  \cos\delta_{\ell}\cdot j_{\ell}(kr)-\sin\delta_{\ell}\cdot y_{\ell}(kr)\right]
Y_{\ell,\ell_{z}}(\theta,\phi). \label{wavefunction}%
\end{equation}
Here $k$ is the relative momentum between the scattering particles,
and $\delta_{\ell}$ is the phase shift for partial wave $\ell$. \ We can extract
$k$ from the energy of the standing wave, and the phase shift $\delta_{\ell}$
is determined by setting the wave function in Eq.~(\ref{wavefunction}) to
zero at the wall boundary.

When the total intrinsic spin of the two nucleons is nonzero,
spin-orbit coupling generates mixing between partial waves. $\ $In this case
the standing wave at the wall boundary is decomposed into spherical harmonics
and coupled-channel equations are solved to extract the phase shifts and
mixing angles.  The spherical wall method
was used to calculate phase shifts and mixing angle for low-energy nucleon-nucleon
scattering \cite{Borasoy:2007vi}.
Recently the spherical wall approach has been improved
in accuracy and computational efficiency \cite{Lu:2015riz}.
In the improved approach one projects onto spherical harmonics
$Y_{\ell,\ell_z}$ with angular momentum quantum numbers $\ell,\ell_z$.  In this manner one constructs radial position states for a given partial wave,
\begin{equation}
|r\rangle^{\ell,\ell_z} = \sum_{{\bf r'}}Y_{\ell,\ell_z}({\bf\hat{r}'})\delta_{r,|{\bf r'}|}|{\bf r'} \rangle.
\end{equation} We require that $r$ is less than half the box length $L/2$.
 Using this technique we are essentially constructing a radial position basis for each partial wave. 

It is also useful to introduce auxiliary potentials in the region lying just in front
of the spherical wall boundary \cite{Lu:2015riz}. The auxiliary potential
is a spherical attractive well that is positioned in front
of the spherical wall boundary.  We can tune to any scattering energy by
adjusting the depth of the well. For systems with partial wave mixing due
to spin-orbit coupling, we also include a Hermitian but imaginary off-diagonal
auxiliary potential
for the two coupled channels.  This breaks time reversal
symmetry, and the resulting standing wave solutions now have both real and imaginary
parts that are linearly independent.  From the real and imaginary solutions
one can determine the scattering phase shifts and mixing angle at any
given value of the scattering energy.

This spherical wall approach has been used together with a technique called the adiabatic projection method to study nuclear scattering and reactions on the lattice.  The adiabiatic projection method \cite{Pine:2013zja,Elhatisari:2014lka,Elhatisari:2015iga,Rokash:2015hra,Elhatisari:2016owd} is a general framework that produces a low-energy effective theory for
clusters of particles which becomes exact in the limit of large projection
time.
For the case of two-cluster scattering, we consider a set of two cluster states
$|{\bf R}\rangle$ labeled by the spatial separation vector {\bf R}. The initial
wave
functions are wave packets which, for large $|{\bf R}|$, factorize into a
product of two
individual clusters,
\begin{equation}
|{\bf R}\rangle=\sum_{{\bf r}} |{\bf r}+{\bf R}\rangle_1\otimes|{\bf r}\rangle_2.
\label{eqn:single_clusters}
\end{equation}
The summation over $\bf {r}$ is required to produce states with 
total momentum equal to zero. We bin the initial cluster states together
according to radial distance and angular momentum. In this manner, we form
radial 
position states with projected angular momentum quantum numbers, which we
label $|R\rangle^{\ell,\ell_z}$. 

The next step is to multiply by powers of the transfer matrix in order to
form ``dressed'' cluster
states. This produces states that approximately span the set of low-energy cluster-cluster scattering
states in our periodic box. We discuss the transfer matrix formalism in detail later in this chapter. After $n_t$ time steps, we have the dressed cluster
states 
\begin{equation}
\vert R\rangle^{\ell,\ell_z}_{n_t} = M^{n_t}|R\rangle^{\ell,\ell_z}.
\end{equation}
These dressed cluster states are then used to compute matrix
elements of the transfer matrix $M$,
\begin{equation}
\left[M_{n_t}\right]^{\ell,\ell_z}_{R',R} =\ ^{\ell,\ell_z}_{\!\!\!\!\!\quad{n_t}}\langle
R'\vert M \vert R\rangle^{\ell,\ell_z}_{n_t}.
\label{Hmatrix}
\end{equation}
Since such states are not orthogonal, we also compute a norm
matrix
\begin{equation}
\left[N_{n_t}\right]^{\ell,\ell_z}_{R',R} =\ ^{\ell,\ell_z}_{\!\!\!\!\!\quad{n_t}}\langle
R'\vert R\rangle^{\ell,\ell_z}_{n_t}. 
\label{eqn:norm}
\end{equation}
The ``radial adiabatic transfer matrix'' is defined as the matrix product
\begin{equation}
\left[ {M^a_{n_t}} \right]^{\ell,\ell_z}_{R',R} = 
\left[N_{n_t}^{-\frac{1}{2}}M_{n_t}
N_{n_t}^{-\frac{1}{2}} \right]^{\ell,\ell_z}_{R',R},
\label{eqn:Adiabatic-Hamiltonian}
\end{equation}
and the scattering phase shifts can then be determined from the standing waves
of the radial adiabatic transfer matrix.  

\section{Lattice formalisms}

\begin{figure}
[pb]
\begin{center}
\includegraphics[width=3in
]%
{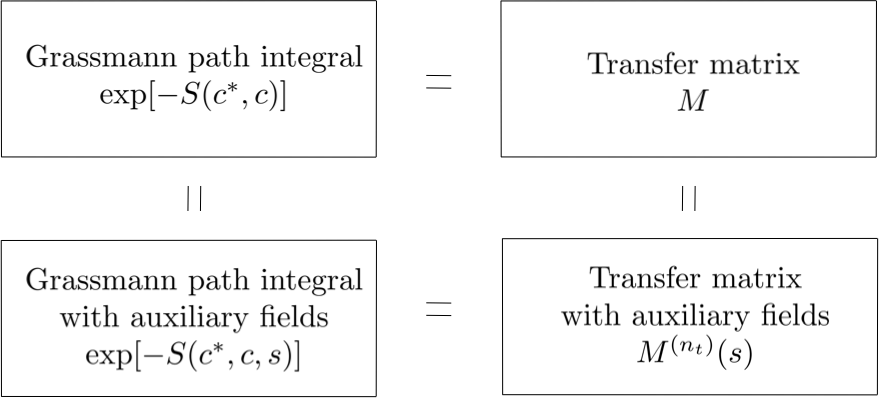}%
\caption{A schematic diagram of the different lattice formulations, namely, the Grassmann path integral, transfer matrix operator, Grassmann
path integral with auxiliary fields, and transfer matrix operator with auxiliary
fields.}%
\label{formalism}%
\end{center}
\end{figure}

Throughout our discussion of the lattice formalism we use dimensionless
parameters and operators corresponding with physical values times
the
appropriate power of the spatial lattice spacing $a$. In our notation the
three-component integer vector ${\bf n}$ labels the lattice sites of a
three-dimensional periodic lattice with dimensions $L^{3}$. The spatial
lattice unit vectors are denoted 
$\mathbf{\hat{l}}$ = $\mathbf{\hat{1}}$,
$\mathbf{\hat{2}}$, $\mathbf{\hat{3}}$.
We use $n_t$ to label lattice steps in the temporal direction, and $L_{t}$
denotes the total number of lattice time steps. \ The temporal lattice spacing
is given by $a_{t}$, and $\alpha_{t}=a_{t}/a$ is the ratio of the temporal
to
spatial lattice spacing. \ We also define $h=\alpha_{t}/(2m)$, where $m$
is
the nucleon mass in lattice units. In Fig.~\ref{formalism} we show a diagram
of the four different but exactly equivalent lattice formulations that we
discuss, the Grassmann path integral, transfer matrix operator, Grassmann
path integral with auxiliary fields, and transfer matrix operator with auxiliary
fields.

\subsection{Grassmann path integral}

We define the lattice action starting from the lattice Grassmann path integral
action without auxiliary fields.
\ This
is the simplest formulation in which to derive the lattice Feynman rules.
 We let $c$ and $c^*$ be anticommuting Grassmann fields for the nucleons.
\ In our notation $c$ is a column vector composed of the spin-isospin nucleon
degrees of freedom $c_i$, while $c^*$ is a row vector of the components
$c^*_i$.  The Grassmann fields are periodic with respect to
the spatial extent of the $L^{3}$ lattice,%
\begin{equation}
c_i({\bf n}+L\hat{1},n_t)=c_i({\bf n}+L\hat{2},n_t)=c_i({\bf n}+L\hat{3},n_t)=c_i({\bf n},n_t),
\end{equation}%
\begin{equation}
c_i^{\ast}({\bf n}+L\hat{1},n_t)=c_i^{\ast}({\bf n}+L\hat{2}%
,n_t)=c_i^{\ast}({\bf n}+L\hat{3},n_t)=c_i^{\ast}({\bf n},n_t),
\end{equation}
and antiperiodic along the temporal direction,%
\begin{equation}
c_i({\bf n},n_t+L_{t})=-c_i({\bf n},n_t),
\end{equation}%
\begin{equation}
c_i^{\ast}({\bf n},n_t+L_{t})=-c_i^{\ast}({\bf n},n_t).
\end{equation}
We write $DcDc^{\ast}$ as shorthand for the integral measure,%
\begin{equation}
DcDc^{\ast}=\prod_{{\bf n},n_t,i}dc_{i}({\bf n}%
,n_t)dc_{i}^{\ast}({\bf n},n_t).
\end{equation}
We use the usual convention for Grassmann integration,%
\begin{equation}
\int dc_{i}({\bf n},n_t)=\int dc_{i}^{\ast}({\bf n},n_t)=0\text{,}%
\end{equation}%
\begin{equation}
\int dc_{i}({\bf n},n_t)c_{i}({\bf n},n_t)=\int dc_{i}^{\ast}(\vec
{n},n_t)c_{i}^{\ast}({\bf n},n_t)=1\text{ \ (no sum on }i\text{)}.
\end{equation}
We consider the Grassmann path integral%
\begin{equation}
\mathcal{Z}=\int DcDc^{\ast}\exp\left[  -S\left(c^{\ast},c\right)  \right]
, \label{defining_Z}%
\end{equation}
where the lattice action can be broken into a free part and interacting part,
\begin{equation}
S(c^*,c)=S_{\text{free}}(c^{\ast},c)+S_\text{int}(c^{\ast},c).
\label{path_nonaux}%
\end{equation}
The free part is the free non-relativistic nucleon action, which is 
\begin{align}
S_{\text{free}}(c^{\ast},c)  &  =\sum_{{\bf n},n_t
}  c_{}^{\ast}({\bf n},n_t) \left[ c({\bf n},n_t+1)-c({\bf n},n_t)\right]
+\alpha_t \sum_{n_t} K^{(n_t)}(c^*,c),
\end{align}
where
\begin{align}
K^{(n_t)}(c^*,c) =\sum_{k=0,1,2,\cdots} (-1)^k \frac{w_k}{2m} \sum_{{\bf
n},{\bf \hat{l}}} c^{\ast}({\bf n},n_t) \left[ c({\bf n}+k{\bf\hat{l}},n_t)
+ c({\bf n}-k{\bf \hat{l}},n_t)\right], 
\end{align}
and the hopping coefficients $w_k$ correspond to a hopping parameter expansion
of the squared momentum,
\begin{align}
P^2({\bf p})=2\sum_{k=0,1,2,\cdots}\sum_{l=1,2,3}%
(-1)^{k}w_{k}\cos\left(kp_{l}\right).
\end{align}
The hopping coefficients are chosen to match the continuum  relation 
\begin{align}
P^2({\bf p})={\bf p}^2,
\end{align}
up to some chosen level of lattice discretization error.  The hopping coefficients
$w_k$ for a few different lattice actions are shown in Table~\ref{hopping_coeff}.

\begin{table}[tbh]
\caption{Hopping coefficients $w_k$ for several lattice actions}%
\label{hopping_coeff}%
\begin{center}
\begin{tabular}{p{2cm}p{3cm}p{3cm}p{3cm}}
\hline\noalign{\smallskip}
coefficient& standard & $O(a^{2})$-improved & $O(a^{4})$-improved \\
\noalign{\smallskip}\svhline\noalign{\smallskip}
$w_{0}$ & $1$ & $5/4$ & $49/36$\\
$w_{1}$ & $1$ & $4/3$ & $3/2$\\
$w_{2}$ & $0$ & $1/12$ & $3/20$\\
$w_{3}$ & $0$ & $0$ & $1/90$\\
\noalign{\smallskip}\hline\noalign{\smallskip}
\end{tabular}
\end{center}
\end{table}

\subsection{Transfer matrix operator}

Let $a_{i}({\bf n})$ and
$a_{i}^{\dagger}({\bf n})$ denote fermion annihilation and creation operators
for the nucleon component $i$ at lattice site ${\bf n}$.  The shorthand $a_{}({\bf
n})$ represents a column vector of nucleon components $a_{i}({\bf n})$, and
$a^{\dagger}({\bf n})$ represents a row vector of components $a^{\dagger}_{i}({\bf
n})$.  We can write any Grassmann path
integral with instantaneous interactions as the trace of a product of
operators using the identity \cite{Creutz:1988wv,Creutz:1999zy}%
\begin{align}
&  {\rm Tr}\left\{  \colon F_{L_{t}-1}\left[  a_{}^{\dagger}({\bf n}^{\prime}),a({\bf
n})\right]  \colon\times\cdots\times\colon
F_{0}\left[  a_{}^{\dagger}({\bf n}^{\prime}),a({\bf n})\right]
\colon\right\} \nonumber\\
&  =\int DcDc^{\ast}\exp\left\{  \sum_{n_t=0}^{L_{t}-1}\sum_{{\bf n},i}%
c_{i}^{\ast}({\bf n},n_t)\left[  c_{i}({\bf n},n_t)-c_{i}({\bf n}%
,n_t+1)\right]  \right\} \nonumber\\
&  \qquad\qquad\qquad\times\prod_{n_t=0}^{L_{t}-1}F_{n_t}\left[
c_{}^{\ast}({\bf n}^{\prime},n_t),c({\bf n},n_t)\right]
 ,
\label{correspondence}%
\end{align}
where $c_{i}({\bf n},L_{t})=-c_{i}({\bf n},0)$.

Let us define the free non-relativistic lattice Hamiltonian
\begin{equation}
H_{\rm free}(a^{\dagger},a) =\sum_{k=0,1,2,\cdots} (-1)^k \frac{w_k}{2m}
\sum_{{\bf n},{\bf \hat{l}}} a^{\dagger}({\bf n}) \left[ a({\bf n}+k{\bf\hat{l}})
+ a({\bf n}-k{\bf \hat{l}})\right].
\end{equation}
We write the interaction term as $H_{\rm int}(a^{\dagger},a)$, so that our
total Hamiltonian is
\begin{equation}
H(a^{\dagger},a) = H_{\rm free}(a^{\dagger},a) + H_{\rm int}(a^{\dagger},a).
\end{equation}Using the correspondence Eq.~(\ref{correspondence}), we can
rewrite the path
integral $\mathcal{Z}$ defined in Eq.~(\ref{defining_Z}) as a transfer-matrix
partition function,%
\begin{equation}
\mathcal{Z}={\rm Tr}\left(  M^{L_{t}}\right)  ,
\end{equation}
where $M$ is the normal-ordered transfer matrix operator%
\begin{equation}
M=:\exp\left[  -H(a^{\dagger},a)\alpha_{t}\right]  :. \label{transfer_noaux}%
\end{equation}
Roughly speaking, the transfer matrix operator is the exponential of the
Hamiltonian operator over one Euclidean lattice time step.
\ In order to satisfy the identity Eq.~(\ref{correspondence}), the exact
definition of the transfer matrix is the normal-ordered exponential as defined
in Eq.~(\ref{transfer_noaux}).

\begin{figure}[ptb]%
\centering
\sidecaption
\includegraphics[
height=7.00cm
]%
{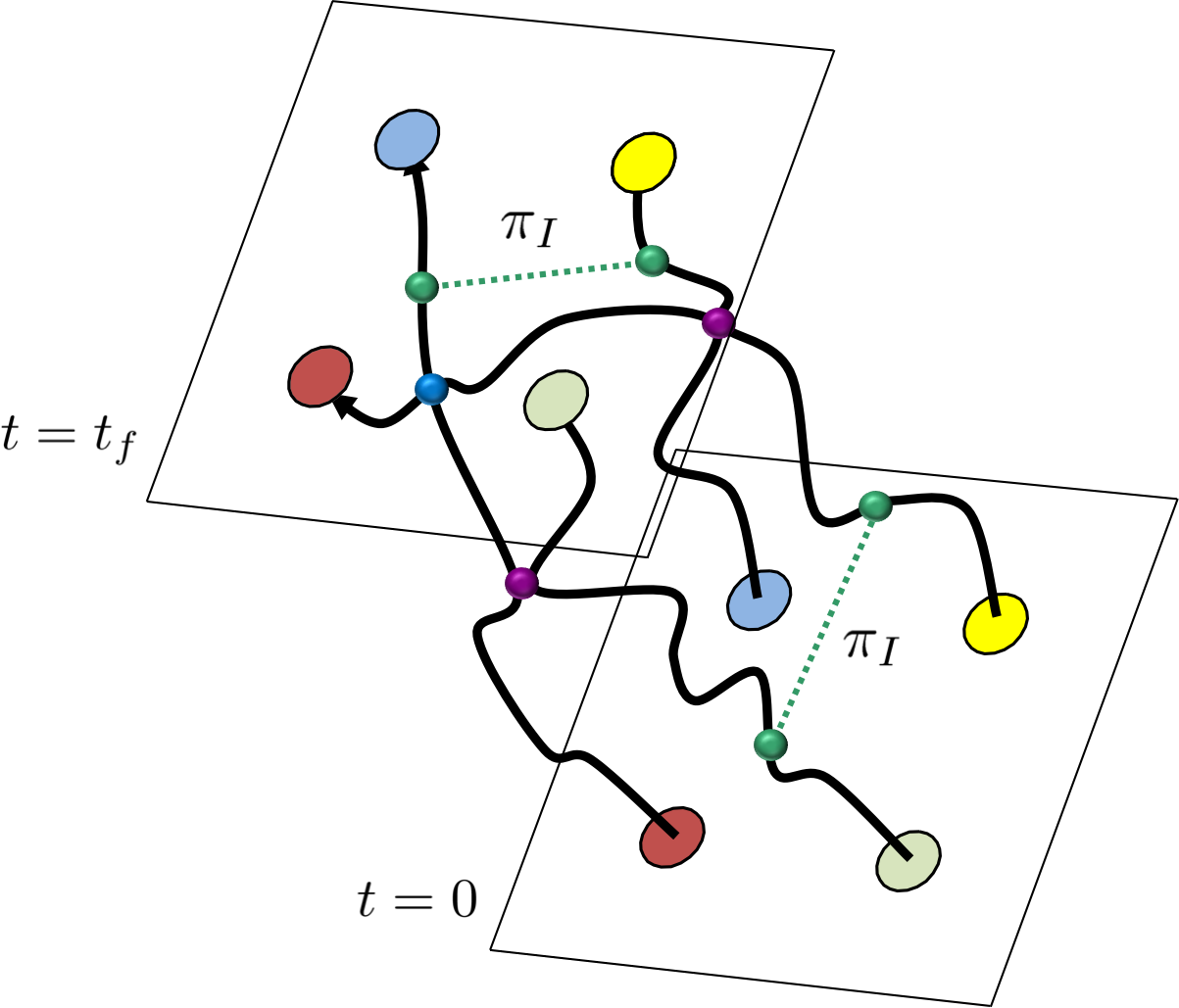}%
\caption{A sketch showing nucleons which evolve with each time step.  At leading order in chiral effective field theory, the interactions include
two contact
interactions and the exchange of pions.}%
\label{worldlines}%
\end{figure}

In this transfer matrix formalism, one can do simulations of nucleons using Monte Carlo, and this would essentially be a lattice version of diffusion or Green's function Monte Carlo \cite{Carlson:2014vla}.  Visually one can view the nucleons as interacting with each other while diffusing in space with each time step, as indicated in Fig.~\ref{worldlines}.  At leading order in chiral effective field theory, the
interactions include two independent $S$-wave contact interactions and the exchange of pions.  We discuss these interactions in detail in the following.

\subsection{Grassmann path integral with auxiliary field}

We assume that there exists an integral relation that allows us to write
$\exp\left[-S_{\rm int}(c^*,c)\right]$ as an integral over auxiliary fields.
 The purpose of the auxiliary field transformation is to decouple the interactions
among the nucleons.  Instead the interactions will be between the nucleons
and the auxiliary fields.  

We illustrate using the interactions that appear at leading order in
chiral effective field theory.  For pedagogical purposes we discuss the simplest possible implementation of the leading order action on the lattice.  We first consider a zero-range contact interaction
which is independent of
nucleon spin and isospin.
The action has the form 
\begin{equation}
S^{C}_\text{int}(c^{\ast},c) = \alpha_{t}\frac{C}{2} \sum_{{\bf n},n_t} \left[c^{\ast}({\bf
n},n_t)c({\bf n},n_t)\right]^2.
\end{equation}
We can write this as
\begin{equation}
\exp\left[-S^C_{\rm int}(c^*,c)\right] = \int Ds \; \exp\left[-S_{ss}(s)
- S_{s}(c^*,c,s)\right]
\label{aux}
\end{equation}
for auxiliary field $s({\bf n},n_t)$, 
where
\begin{align}
S_{ss}(s) = \frac{1}{2} & \sum_{{\bf n},n_t} s^{2}({\bf n},n_t),
 \\
S_{s}(c^*,c,s) =  \sqrt{-C\alpha_{t}} & \sum_{{\bf n},n_t} s({\bf n},n_t)c^{\ast}({\bf
n},n_t)c({\bf n},n_t).  
\end{align}
In our definition of the integration measure $Ds$, we include a factor of
$1/\sqrt{2\pi}$ for each degree of freedom.

Next we consider an isospin-dependent contact interaction 
\begin{equation}
S^{C'}_\text{int}(c^{\ast},c) = \alpha_{t}\frac{C'}{2} \sum_{{\bf n},n_t,I}
\left[c^{\ast}({\bf
n},n_t)\tau_I c({\bf n},n_t)\right]^2,
\end{equation}
where $\tau_I$ for $I=1,2,3$ are the Pauli matrices in isospin space.  Then
we can use
\begin{equation}
\exp\left[-S^{C'}_{\rm int}(c^*,c)\right] = \int \prod_I Ds_I \exp\left[-S_{s_Is_I}(s_I)
- S_{s_I}(c^*,c,s_{I})\right]
\label{aux2}
\end{equation}
for auxiliary fields $s_I({\bf n},n_t)$ where
\begin{align}
S_{s_Is_I}(s_I)=\frac{1}{2} & \sum_{{\bf n},n_t,I} s_I^{2}({\bf n},n_t),
\\
S_{s_I}(c^*,c,s_{I}) = \sqrt{-C'\alpha_{t}} & \sum_{{\bf n},n_t,I} s_{I}({\bf
n},n_t)c^{\ast}({\bf
n},n_t)\tau_Ic({\bf n},n_t).  
\end{align}

Finally we work with the one-pion exchange potential (OPEP).  In this case
the pion acts much like the auxiliary fields.  However there are also spatial
correlations in the quadratic part of the pion action and a gradient coupling
between the pions and nucleons.
 The one-pion exchange interaction on the lattice can written as
\begin{equation}
\exp\left[-S^{\rm OPEP}_{\rm int}(c^*,c)\right] = \int \prod_I D\pi_I
\exp\left[-S_{\pi_I\pi_I}(\pi_I) - S_{\pi_I}(c^*,c,\pi_{I})\right].
\end{equation}
The free pion action is
\begin{align}
S_{\pi_I\pi_I}(\pi_I)= & \frac{1}{2}\alpha_{t}m^2_{\pi} \sum_{{\bf n},n_t,I}\pi^2_{I}({\bf
n},n_t) \\
& +\frac{1}{2}\alpha_{t}%
\sum_{k=0,1,2,\cdots}(-1)^k w_k \sum_{{\bf n},n_t,I,{\bf \hat{l}}} \pi_{I}({\bf
n},n_t)
\left[ \pi_{I}({\bf n}+k{\bf \hat{l}},n_t) + \pi_{I}({\bf n}-k{\bf \hat{l}},n_t)
\right],
\end{align}
with the coefficient $w_k$ as defined in Table~\ref{hopping_coeff} and $m_{\pi}$ is the pion mass.  At leading order we do not
consider any isospin-breaking effects.  The pion coupling to the nucleon is\begin{align}
S_{\pi_I}(c^*,c,\pi_{I}) = \frac{g_A \alpha_t}{2 f_{\pi}} \sum_{{\bf n},n_t,l,I}
\Delta_k \pi_{I}({\bf
n},n_t)c^{\ast}({\bf
n},n_t)\sigma_k\tau_Ic({\bf n},n_t),  
\end{align}
where $\sigma_l$ for $l=1,2,3$ are the Pauli matrices in spin space and 
\begin{equation}
\Delta_l \pi_{I}({\bf
n},n_t)=\frac{1}{2}\sum_{k=1,2,\cdots} (-1)^{k-1} o_k \left[  \pi_{I}({\bf
n}+k{\bf \hat{l}},n_t) - \pi_{I}({\bf
n}-k{\bf \hat{l}},n_t)\right],
\end{equation}
with coefficients $o_k$ corresponding to a hopping parameter expansion of the
momentum,
\begin{align}
P(p_l)=\sum_{k=1,2,\cdots}%
(-1)^{k-1} o_{k}\sin\left(kp_{l}\right).
\end{align}
Here
$g_A$ is the axial-vector coupling constant, and $f_{\pi}$ is the pion
decay constant.  The hopping coefficients can be chosen to match the continuum result
\begin{equation}
P(p_l)= p_l.
\end{equation}
The hopping coefficients $o_k$ for a few different lattice actions are shown
in Table~\ref{hopping_coeff_2}.
\begin{table}[tbh]
\caption{Hopping coefficients $o_k$ for several lattice actions.}%
\begin{center}
\label{hopping_coeff_2}%

\begin{tabular}{p{2cm}p{3cm}p{3cm}p{3cm}}
\hline\noalign{\smallskip}
coefficient& standard & $O(a^{2})$-improved & $O(a^{4})$-improved \\
\noalign{\smallskip}\svhline\noalign{\smallskip}
$o_{1}$ & $1$ & $4/3$ & $3/2$\\
$o_{2}$ & $0$ & $1/6$ & $3/10$\\
$o_{3}$ & $0$ & $0$ & $1/30$ \\
\noalign{\smallskip}\hline\noalign{\smallskip}
\end{tabular}
\end{center}
\end{table}

\subsection{Transfer matrix operator with auxiliary field}
Using the equivalence in Eq.~(\ref{correspondence}), we can write $\mathcal{Z}$
as the trace of a product of transfer matrix operators which depend on the
auxiliary field,
\begin{equation}
\mathcal{Z} = \int Ds \prod_I \left(Ds_I D\pi_I\right) \; 
\exp{\left[-S_{ss}(s)
-S_{s_Is_I}(s_I)
-S_{\pi_I\pi_I}(\pi_I)\right]}
 {\rm Tr}\left\{ M^{(L_t-1)}\cdots M^{(0)}\right\}.
\end{equation}
The transfer matrix at time step $n_t$ is given by
\begin{equation}
M^{(n_t)}=\colon\exp\left[ -H^{(n_t)}(a^{\dagger},a,s,s_I,\pi_I)\alpha_{t}
\right]
\colon,
\end{equation}
where
\begin{equation}
H^{(n_t)}(a^{\dagger},a,s,s_I,\pi_I)\alpha_{t}=H_{\text{free}}(a^{\dagger},a)\alpha_{t}+S^{(n_t)}_s(a^{\dagger},a,s)+S^{(n_t)}_{s_I}(a^{\dagger},a,s_{I})+S^{(n_t)}_{\pi_I}(a^{\dagger},a,\pi_{I}),
\end{equation}
and
\begin{equation}
S^{(n_t)}_{s}(a^{\dagger},a,s) =  \sqrt{-C\alpha_{t}} \sum_{\bf n} s({\bf
n},n_t)a^{\dagger}({\bf
n})a({\bf n}),  
\end{equation}
\begin{equation}
S^{(n_t)}_{s_I}(a^{\dagger},a,s_{I}) = \sqrt{-C'\alpha_{t}} \sum_{{\bf n},I}
s_{I}({\bf
n},n_t)a^{\dagger}({\bf
n})\tau_Ia({\bf n}),
\end{equation}
\begin{equation}
S^{(n_t)}_{\pi_I}(a^{\dagger},a,\pi_{I}) = \frac{g_A \alpha_t}{2 f_{\pi}}
\sum_{{\bf n},k,I}
\Delta_k \pi_{I}({\bf
n},n_t)a^{\dagger}({\bf
n})\sigma_k\tau_I a({\bf n}).
\end{equation}

\section{Projection Monte Carlo}
Let us consider a system with $A$ nucleons.  We can create a general single-nucleon
state using 
creation operators acting on the vacuum with coefficient function $f({\bf
n})$. We write $f({\bf n})$ as a column vector in the space of nucleon spin and isospin components, and the single-nucleon
state can be written as\begin{equation}
\left| f \right> = \sum_{\bf n}a^{\dagger}({\bf n})f({\bf n}) \left| 0 \right>.
\end{equation}
For our projection Monte Carlo calculation we take our A-body initial
state
to be a Slater determinant of single nucleon states,
\begin{equation}
\left| f_1,\cdots ,f_A \right> =   \left[\sum_{\bf n}a^{\dagger}({\bf n})f_1({\bf
n}) \right] \cdots \left[\sum_{\bf n}a^{\dagger}({\bf n})f_A({\bf n}) \right]
\left| 0 \right>.
\end{equation} We use the same construction for the $A$-body final state.

For the purposes of coding the projection Monte Carlo calculation, it is
convenient to view the identical nucleons as having a hidden index $j=1,\cdots
,A$ that makes all of the nucleons distinguishable.  If we antisymmetrize
all physical states over this extra index then all physical observables are
exactly recovered.   So our initial state $\left| f_1,\cdots ,f_A \right>$
becomes
\begin{align}
 \frac{1}{\sqrt{A!}} & \sum_{P}   \left[\sum_{\bf n}a^{\dagger}_{[P(1)]}({\bf n})f_1({\bf
n}) \right] \cdots \left[\sum_{\bf n}a^{\dagger}_{[P(A)]}({\bf n})f_A({\bf
n}) \right]
\left| 0 \right> \nonumber \\
 & =\frac{1}{\sqrt{A!}} \sum_{P'} {\rm sgn}(P') \left[\sum_{\bf n}a^{\dagger}_{[1]}({\bf
n})f_{P'(1)}({\bf
n}) \right] \cdots \left[\sum_{\bf n}a^{\dagger}_{[A]}({\bf n})f_{P'(A)}({\bf
n}) \right]
\left| 0 \right>,
\end{align}
where the summations are over all permutations $P$, and ${\rm sgn}$ is
the sign of the permutation.
With these hidden indices our normal-ordered auxiliary-field transfer matrix
$M^{(n_t)}$  becomes 
\begin{equation}
\left[1-H^{(n_t)}(a_{[1]}^{\dagger},a_{[1]},s,s_I,\pi_I)\alpha_{t}\right]
\cdots \left[1-H^{(n_t)}(a^{\dagger}_{[A]},a_{[A]},s,s_I,\pi_I)\alpha_{t}\right]
\end{equation}
We see that the higher-order powers of the exponential vanish due to normal ordering.

In the projection Monte Carlo calculation we compute the amplitude
\begin{equation}
Z(n_t) = \left< f_1,\cdots ,f_A \right|  M^{(n_t-1)} \cdots M^{(0)} \left|
f_1,\cdots ,f_A \right>
\end{equation}
for $n_t=L_t$ and $n_t=L_t-1$.  In the limit of large $L_t$ the amplitudes will be dominated
by the state with the lowest energy $E_0$ and nonzero overlap with $\left|
f_1,\cdots ,f_A \right>$.  In this limit the ratio $Z(n_t)/Z(n_t-1)$
will converge to $\exp(-E_0 \alpha_t)$ from above.

\begin{figure}[ptb]%
\centering
\sidecaption
\includegraphics[
height=7.00cm
]
{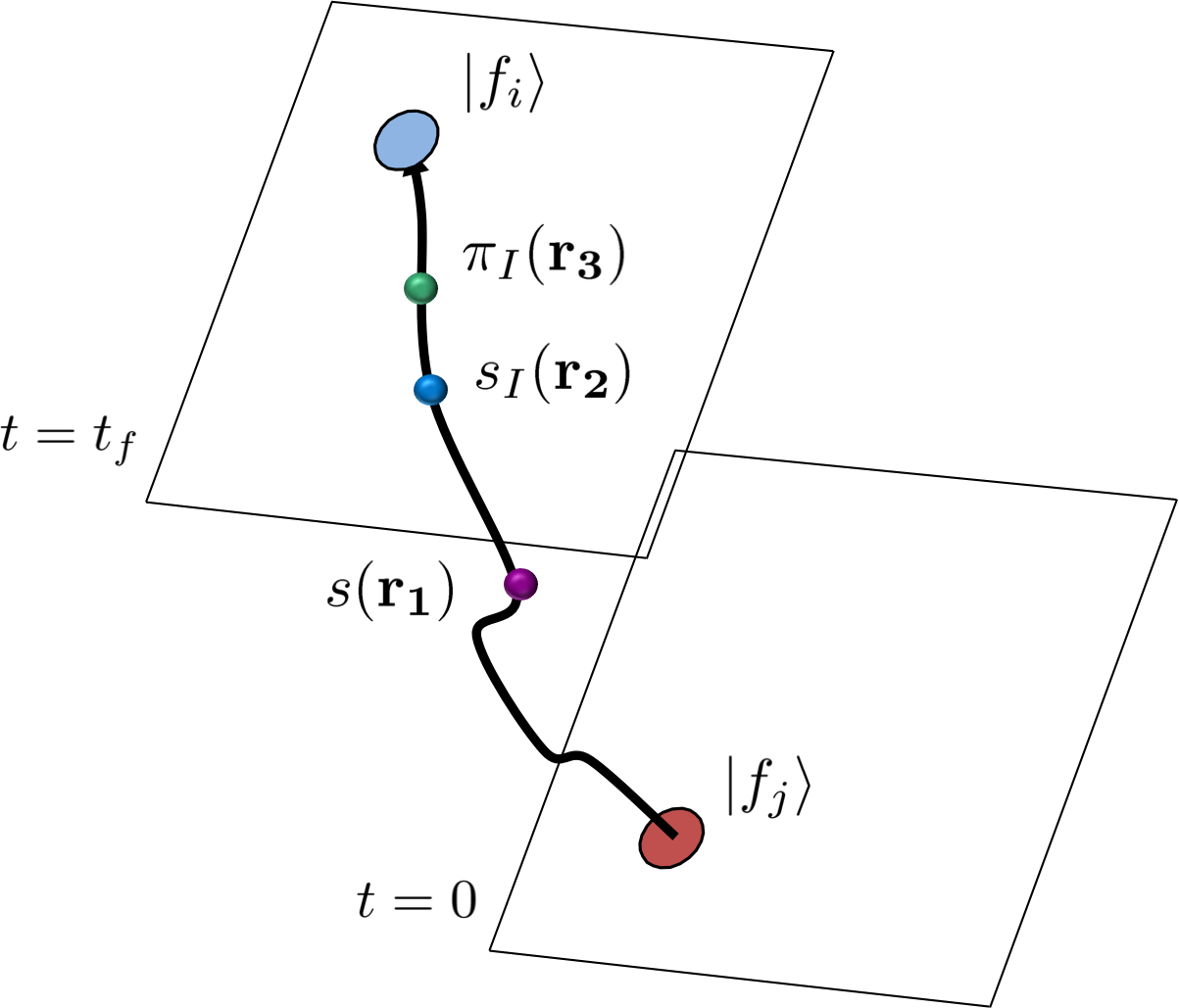}%
\caption{A sketch showing the worldline for a single nucleon with a background
of pion fields and
auxiliary fields.}%
\label{hsworldlines}%
\end{figure}

Each nucleon evolves as a
particle in a fluctuating background of auxiliary fields and pion
fields. \  The original interactions are reproduced after integrating over
the fluctuating auxiliary
and pion fields. For a simulation with $A$ nucleons, the amplitude for
a given configuration of pion and auxiliary fields is proportional to the
determinant of an $A\times A$ matrix ${\bf M}$. \ The
entries of ${\bf M}_{ij}$ are single nucleon worldline amplitudes for a nucleon
starting at state $\left| f_j \right> $ at $t=0$ and ending at state $\left|
f_i \right> $ at $t=t_{f}=L_t\alpha_t$. This is shown in Fig. \ref{hsworldlines}.

In the following we show sample code \ref{label_1} in the Fortran programming language which calculates the 
auxiliary-field transfer matrix multiplications on the left starting from the 
single-nucleon initial states.  We show only the terms which arise from 
the free-nucleon transfer matrix and the auxiliary field $s$.  

\begin{lstlisting}[language=Fortran,caption=Sample code calculating 
the auxiliary-field transfer matrix multiplications on the left starting from the single-nucleon initial states,label={label_1}]
DO nt = nt1+1, nt2     
  DO np = 0,num-1        
    DO nz = 0,L-1; DO ny = 0,L-1; DO nx = 0,L-1; DO ni = 0,1; DO ns = 0,1
              
     zvecs(nx,ny,nz,nt,ns,ni,np) = zvecs(nx,ny,nz,nt-1,ns,ni,np) &
        * (1.D0-6.D0*w0_N*h+CDSQRT(-c0*atovera*(1.D0,0.D0))*s(nx,ny,nz,nt-1))
              
      zvecs(nx,ny,nz,nt,ns,ni,np) = zvecs(nx,ny,nz,nt,ns,ni,np) &
        + w1_N*h*zvecs(MOD(nx+1,L),ny,nz,nt-1,ns,ni,np) &
        + w1_N*h*zvecs(MOD(nx-1+L,L),ny,nz,nt-1,ns,ni,np) &
        + w1_N*h*zvecs(nx,MOD(ny+1,L),nz,nt-1,ns,ni,np) &
        + w1_N*h*zvecs(nx,MOD(ny-1+L,L),nz,nt-1,ns,ni,np) &
        + w1_N*h*zvecs(nx,ny,MOD(nz+1,L),nt-1,ns,ni,np) &
        + w1_N*h*zvecs(nx,ny,MOD(nz-1+L,L),nt-1,ns,ni,np)
              
      IF (improveN >= 1) THEN                 
        zvecs(nx,ny,nz,nt,ns,ni,np) = zvecs(nx,ny,nz,nt,ns,ni,np) & 
          - w2_N*h*zvecs(MOD(nx+2,L),ny,nz,nt-1,ns,ni,np) &
          - w2_N*h*zvecs(MOD(nx-2+L,L),ny,nz,nt-1,ns,ni,np) &
          - w2_N*h*zvecs(nx,MOD(ny+2,L),nz,nt-1,ns,ni,np) &
          - w2_N*h*zvecs(nx,MOD(ny-2+L,L),nz,nt-1,ns,ni,np) &
          - w2_N*h*zvecs(nx,ny,MOD(nz+2,L),nt-1,ns,ni,np) &
          - w2_N*h*zvecs(nx,ny,MOD(nz-2+L,L),nt-1,ns,ni,np)
      END IF
              
      IF (improveN == 2) THEN
        zvecs(nx,ny,nz,nt,ns,ni,np) = zvecs(nx,ny,nz,nt,ns,ni,np) &
          + w3_N*h*zvecs(MOD(nx+3,L),ny,nz,nt-1,ns,ni,np) &
          + w3_N*h*zvecs(MOD(nx-3+L,L),ny,nz,nt-1,ns,ni,np) &
          + w3_N*h*zvecs(nx,MOD(ny+3,L),nz,nt-1,ns,ni,np) &
          + w3_N*h*zvecs(nx,MOD(ny-3+L,L),nz,nt-1,ns,ni,np) &           
          + w3_N*h*zvecs(nx,ny,MOD(nz+3,L),nt-1,ns,ni,np) &
          + w3_N*h*zvecs(nx,ny,MOD(nz-3+L,L),nt-1,ns,ni,np)
      END IF
              
    END DO; END DO; END DO; END DO; END DO
  END DO
END DO

\end{lstlisting}

Similarly, we now show sample code \ref{label_2} which calculates 
the auxiliary-field transfer matrix multiplications on the right starting from 
the single-nucleon final states.  Again we present only the terms arising from the free-nucleon transfer matrix and the auxiliary field $s$.  

\begin{lstlisting}[language=Fortran,caption=Sample code calculating 
the auxiliary-field transfer matrix multiplications on the right starting
from the single-nucleon final states,label={label_2}]
DO nt = nt2,nt1+1,-1
  DO np = 0,num-1   
    DO nz = 0,L-1; DO ny = 0,L-1; DO nx = 0,L-1; DO ni = 0,1; DO ns = 0,1
               
      zdualvecs(nx,ny,nz,nt-1,ns,ni,np) &
        = zdualvecs(nx,ny,nz,nt,ns,ni,np) &
        * (1.D0-6.D0*w0_N*h+CDSQRT(-c0*atovera*(1.D0,0.D0))*s(nx,ny,nz,nt-1))
               
      zdualvecs(nx,ny,nz,nt-1,ns,ni,np) &
        = zdualvecs(nx,ny,nz,nt-1,ns,ni,np) & 
        + w1_N*h*zdualvecs(MOD(nx+1,L),ny,nz,nt,ns,ni,np) &
        + w1_N*h*zdualvecs(MOD(nx-1+L,L),ny,nz,nt,ns,ni,np) &
        + w1_N*h*zdualvecs(nx,MOD(ny+1,L),nz,nt,ns,ni,np) &
        + w1_N*h*zdualvecs(nx,MOD(ny-1+L,L),nz,nt,ns,ni,np) &
        + w1_N*h*zdualvecs(nx,ny,MOD(nz+1,L),nt,ns,ni,np) &
        + w1_N*h*zdualvecs(nx,ny,MOD(nz-1+L,L),nt,ns,ni,np) 
               
      IF (improveN >= 1) THEN
        zdualvecs(nx,ny,nz,nt-1,ns,ni,np) &
          = zdualvecs(nx,ny,nz,nt-1,ns,ni,np) &
          - w2_N*h*zdualvecs(MOD(nx+2,L),ny,nz,nt,ns,ni,np) &
          - w2_N*h*zdualvecs(MOD(nx-2+L,L),ny,nz,nt,ns,ni,np) &
          - w2_N*h*zdualvecs(nx,MOD(ny+2,L),nz,nt,ns,ni,np) &
          - w2_N*h*zdualvecs(nx,MOD(ny-2+L,L),nz,nt,ns,ni,np) &
          - w2_N*h*zdualvecs(nx,ny,MOD(nz+2,L),nt,ns,ni,np) &
          - w2_N*h*zdualvecs(nx,ny,MOD(nz-2+L,L),nt,ns,ni,np) 
      END IF
               
      IF (improveN == 2) THEN
        zdualvecs(nx,ny,nz,nt-1,ns,ni,np) &
        = zdualvecs(nx,ny,nz,nt-1,ns,ni,np) &
        + w3_N*h*zdualvecs(MOD(nx+3,L),ny,nz,nt,ns,ni,np) &
        + w3_N*h*zdualvecs(MOD(nx-3+L,L),ny,nz,nt,ns,ni,np) &
        + w3_N*h*zdualvecs(nx,MOD(ny+3,L),nz,nt,ns,ni,np) &
        + w3_N*h*zdualvecs(nx,MOD(ny-3+L,L),nz,nt,ns,ni,np) &
        + w3_N*h*zdualvecs(nx,ny,MOD(nz+3,L),nt,ns,ni,np) &
        + w3_N*h*zdualvecs(nx,ny,MOD(nz-3+L,L),nt,ns,ni,np) 
      END IF
               
    END DO; END DO
  END DO; END DO; END DO
END DO

\end{lstlisting}

In the following we show sample code \ref{label_3} where these transfer matrix product multiplications are called as subroutines and used to compute the determinant and inverse of the matrix of single-nucleon amplitudes {\bf M}.  

\begin{lstlisting}[language=Fortran,caption=Sample code where transfer matrix product multiplications are called and used to compute the determinant and inverse of the matrix of single-nucleon
amplitudes.,label={label_3}]
CALL getzvecs(s,sI,zvecs,zwave,Lt,0,pion,ztau2x2,n_f)            
CALL getzdualvecs(s,sI,zdualvecs,zdualwave,Lt,0,pion,ztau2x2,n_f)
CALL getinvcorr(zvecs,zdualvecs,zldeter,zcorrmatrix,zcorrinv,Lt)

aldeterabs = DBLE(zldeter)
zdeterphase = CDEXP((0.D0,1.D0)*DIMAG(zldeter))
act = bose - aldeterabs
\end{lstlisting}

\section{Importance sampling}

We do importance sampling according to the positive measure 
\begin{equation}
|Z(L_t)|\exp{\left[-S_{ss}(s)
-S_{s_Is_I}(s_I)
-S_{\pi_I\pi_I}(\pi_I)\right]},
\end{equation} 
and use hybrid Monte Carlo to do global updates of the auxiliary and pion
fields. The hybrid Monte Carlo~(HMC) algorithm
\cite{Scalettar:1986uy,Gottlieb:1987mq,Duane:1987de} is efficient in quickly generating decorrelated configurations for
each
auxiliary and pion field.  Here we describe the
updating algorithm for the $s$ field. The updating of the $s_I$ and $\pi_I$
fields proceed in a very similar fashion.  In
general terms, the HMC algorithm can be described by means of a probability
weight $P(s)$
\begin{equation}
P(s)\propto\exp[-V(s)],
\end{equation}
where $V(s)$ is in general a non-local function of the field $s({{\bf n}%
,n_t}),$ and a molecular dynamics~(MD) Hamiltonian,
\begin{equation}
\quad\quad H(s,p)\equiv\frac{1}{2}\sum_{{\bf n},n_t%
}\left[p_s({{\bf n},n_t})\right]  ^{2}+V(s).
\end{equation}
Classical Hamiltonian dynamics is introduced by defining
the
momentum $p_s({{\bf n},n_t})$ conjugate to $s({{\bf n},n_t})$.

Given an arbitrary initial configuration $s^{0}({{\bf n},n_t})$, the
conjugate momentum is chosen from a random Gaussian distribution according
to
\begin{equation}
P[p_s^{0}({\bf n},n_t)]\propto\exp\left\{  -\frac{1}{2}\left[  p_s^{0}({\bf
n},n_t)\right]  ^{2}\right\}  , \label{Gdistr}%
\end{equation}
after which the Hamiltonian equations of motion are integrated numerically
with a small but nonzero step size $\varepsilon_{\mathrm{step}}$.  This method begins
with a
\textquotedblleft half-step\textquotedblright\ forward in the conjugate
momentum,
\begin{equation}
\tilde{p}_s^{0}({\bf n},n_t)=p_s^{0}({\bf n},n_t)-\frac{\varepsilon
_{\mathrm{step}}}{2}\left[  \frac{\partial V(s)}{\partial s({\bf n},n_t%
)}\right]  _{s=s^{0}},
\end{equation}
followed by repeated updates of $s$ and $\tilde{p}_s$ according to
\begin{equation}
s^{i+1}({\bf n},n_t)=s^{i}({\bf n},n_t)+\varepsilon_{\mathrm{step}}%
\tilde{p}^{i}_s({\bf n},n_t),\quad\quad\tilde{p}^{i+1}_s({\bf n},n_t%
)=\tilde{p}^{i}_s({\bf n},n_t)-\varepsilon_{\mathrm{step}}\left[
\frac{\partial V(s)}{\partial s({\bf n},n_t)}\right]  _{s=s^{i+1}},
\end{equation}
for a specified number of steps $N_{\mathrm{step}}$. This is followed by an additional half-step
backward in $\tilde{p}_s$  given by
\begin{equation}
p_s^{N_{\mathrm{step}}}({\bf n},n_t)=\tilde{p}_s^{N_{\mathrm{step}}}(\vec
{n},n_t)+\frac{\varepsilon_{\mathrm{step}}}{2}\left[  \frac{\partial
V(s)}{\partial s({\bf n},n_t)}\right]  _{s=s^{0}}.
\end{equation}

For algorithmic efficiency the length of such an MD \textquotedblleft trajectory\textquotedblright%
\ should be taken large enough to ensure decorrelation between
successive configurations of the auxiliary field. The evolved configuration
is
then subjected to a \textquotedblleft Metropolis test\textquotedblright%
\ against a random number $r\in\lbrack0,1)$.  The new configuration is accepted if
\begin{equation}
r<\exp\left[  -H(s^{N_{\mathrm{step}}},p^{N_{\rm step }}_s)+H(s^{0}%
,p_s^{0})\right]  .
\end{equation}
It
should be noted that although $H$ is in principle conserved in the MD
evolution, the truncation error of the leapfrog method introduces a systematic
error. The Metropolis test eliminates the need for extrapolation in
$\varepsilon_{\mathrm{step}}$. 

In our case $\exp[-V(s)]$ has the form 
\begin{equation}
|Z(L_t)|\exp{\left[-S_{ss}(s)
-S_{s_Is_I}(s_I)
-S_{\pi_I\pi_I}(\pi_I)\right]},
\end{equation} 
where $Z(L_t)$ is the determinant
of an $A\times A$ matrix of single-nucleon amplitudes {\bf M}.  The derivative of
$V_{}$ is then computed using%
\begin{align}
\frac{\partial V_{}(s)}{\partial s({\bf n},n_t)}  &  =\frac{\partial
S_{ss}(s)}{\partial s({\bf n},n_t)}- \frac{\partial {\rm Re} \left[
\ln \left( \det\mathbf{M} \right) \right]}{\partial s(\vec
{n},n_t)} \nonumber\\
  &  =\frac{\partial
S_{ss}(s)}{\partial s({\bf n},n_t)}- {\rm Re} \left[ \frac{1}{\det\mathbf{M}}\sum
_{k,l}\frac{\partial\det\mathbf{M}}{\partial \mathbf{M}
_{kl}}\frac{\partial  \mathbf{M}_{kl}}{\partial s(\vec
{n},n_t)} \right] \nonumber\\
&  =\frac{\partial S_{ss}(s)}{\partial s({\bf n},n_t)}- {\rm Re} \left[
\sum_{k,l}
\mathbf{M}^{-1}  _{lk}\frac{\partial  \mathbf{M}
_{kl}}{\partial s({\bf n},n_t)} \right].
\end{align}

In the following we show sample code \ref{label_4} calculating the
quadratic part of the action due to the auxiliary fields and pion fields,
\begin{equation}
\frac{1}{2}\sum_{{\bf n},n_t}\left[p_s({{\bf n},n_t})\right]^{2}+\frac{1}{2}\sum_{{\bf n},n_t,I}\left[  p_{s_I}({{\bf n},n_t})\right]^{2}+\frac{1}{2}\sum_{{\bf n},n_t,I}\left[p_{\pi_I}({{\bf n},n_t})\right]^{2}+S_{ss}(s) + S_{s_Is_I}(s_I) + S_{\pi_I\pi_I}(\pi_I).
\end{equation}
In the code we have found it convenient to rescale the pion field by a factor of $\sqrt{q_{\pi}}$ where 
\begin{equation}
q_{\pi}=\alpha_{t} \left( m_{\pi}^2 + 6w_0 \right).
\end{equation}

\begin{lstlisting}[language=Fortran,caption=Sample code calculating the
quadratic part of the action due to the auxiliary fields and pion fields.,label={label_4}]
bose = 0.D0
DO nt = 0,Lt-1
  DO nz = 0,L-1; DO ny = 0,L-1; DO nx = 0,L-1
    bose = bose &
      + s(nx,ny,nz,nt)**2.D0/2.D0 &
      + p_s(nx,ny,nz,nt)**2.D0/2.D0
    DO iso = 1,3
      bose = bose &
        + sI(nx,ny,nz,nt,iso)**2.D0/2.D0 &
        + p_sI(nx,ny,nz,nt,iso)**2.D0/2.D0 &
        + pion(nx,ny,nz,nt,iso)**2.D0/2.D0 &
        + atovera/qpi3*pion(nx,ny,nz,nt,iso)*( &
        - w1_P*pion(MOD(nx+1,L),ny,nz,nt,iso) &
        - w1_P*pion(nx,MOD(ny+1,L),nz,nt,iso) &
        - w1_P*pion(nx,ny,MOD(nz+1,L),nt,iso) &
        + w2_P*pion(MOD(nx+2,L),ny,nz,nt,iso) &
        + w2_P*pion(nx,MOD(ny+2,L),nz,nt,iso) &
        + w2_P*pion(nx,ny,MOD(nz+2,L),nt,iso) &
        - w3_P*pion(MOD(nx+3,L),ny,nz,nt,iso) &
        - w3_P*pion(nx,MOD(ny+3,L),nz,nt,iso) &
        - w3_P*pion(nx,ny,MOD(nz+3,L),nt,iso)) &
        + p_pion(nx,ny,nz,nt,iso)**2.D0/2.D0
    END DO
  END DO; END DO; END DO
END DO

\end{lstlisting}     

In following we show sample code \ref{label_5} which calculates 
\begin{equation}
\left[  \frac{\partial V(s)}{\partial s({\bf n},n_t%
)}\right]  _{s=s^{0}}
\end{equation}
and uses it to compute the
 half-step\ forward in the conjugate
momentum,
\begin{equation}
\tilde{p}_s^{0}({\bf n},n_t)=p_s^{0}({\bf n},n_t)-\frac{\varepsilon
_{\mathrm{step}}}{2}\left[  \frac{\partial V(s)}{\partial s({\bf n},n_t%
)}\right]  _{s=s^{0}}.
\end{equation}

\begin{lstlisting}[language=Fortran,caption=Sample code computing derivative with respect to the auxiliary field and half-step forward in the conjugate momentum.,label={label_5}]

DO npart1 = 0,n_f-1; DO npart2 = 0,n_f-1
  zdcorrmatrix(npart2,npart1) = 0.D0
  DO ni = 0,1; DO ns = 0,1
    zdcorrmatrix(npart2,npart1) = &
      zdcorrmatrix(npart2,npart1) + &
      zdualvecs(nx,ny,nz,nt+1,ns,ni,npart2) &
      *zvecs(nx,ny,nz,nt,ns,ni,npart1) &
      *CDSQRT(-c0*atovera*(1.D0,0.D0))/L**3
  END DO; END DO
END DO; END DO
           
dVds(nx,ny,nz,nt) = s(nx,ny,nz,nt) 

DO npart1 = 0,n_f-1; DO npart2 = 0,n_f-1
  dVds(nx,ny,nz,nt) = dVds(nx,ny,nz,nt) &
    - DBLE(zdcorrmatrix(npart2,npart1) &
    *zcorrinv(npart1,npart2))
END DO; END DO
                 
p_sHMC(nx,ny,nz,nt,0) = &
  p_s(nx,ny,nz,nt) - 0.5D0*eHMC*dVds(nx,ny,nz,nt)
  
\end{lstlisting}

In following code \ref{label_6} we show an example code which performs  the Metropolis test%
\ against a random number $r\in\lbrack0,1)$, with the new configuration being accepted if
\begin{equation}
r<\exp\left[  -H(s^{N_{\mathrm{step}}},p^{N_{\rm step }}_s)+H(s^{0}%
,p_s^{0})\right].
\end{equation}

\begin{lstlisting}[language=Fortran,caption=Sample code which performs  the Metropolis acceptance test,label={label_6}]
IF (ntrial .eq. 1 .or. grnd() .lt. DEXP(-actnew+act)) THEN

  accept = accept + 1.

  DO nt = 0,Lt-1
    DO nz = 0,L-1; DO ny = 0,L-1; DO nx = 0,L-1      
      s(nx,ny,nz,nt) = snew(nx,ny,nz,nt)
    END DO; END DO; END DO
  END DO
  
  DO nt = 0,Lt-1
    DO nz = 0,L-1; DO ny = 0,L-1; DO nx = 0,L-1      
      DO iso = 1,3
        sI(nx,ny,nz,nt,iso) = sInew(nx,ny,nz,nt,iso)
        pion(nx,ny,nz,nt,iso) = pionnew(nx,ny,nz,nt,iso)
      END DO
    END DO; END DO; END DO
  END DO
  
  aldeterabs = aldeternewabs
  zdeterphase = zdeternewphase
            
END IF
\end{lstlisting}

Although the Monte Carlo importance sampling uses only the absolute value of the amplitude, the complex phase
of the amplitude is treated as an observable and is collected  with each
configuration of the auxiliary and pion fields.

\section{Exercises}

\begin{prob}
\label{prob6.1}
Write a lattice hybrid Monte Carlo code which performs updates of the lattice action according to only the quadratic part of the action due to the auxiliary fields and pions,
\begin{equation}
\frac{1}{2}\sum_{{\bf n},n_t}\left[p_s({{\bf n},n_t})\right]^{2}+\frac{1}{2}\sum_{{\bf
n},n_t,I}\left[  p_{s_I}({{\bf n},n_t})\right]^{2}+\frac{1}{2}\sum_{{\bf
n},n_t,I}\left[p_{\pi_I}({{\bf n},n_t})\right]^{2}+S_{ss}(s) + S_{s_Is_I}(s_I)
+ S_{\pi_I\pi_I}(\pi_I).
\end{equation} 
Verify that the change in the action produced by the hybrid Monte Carlo update is scaling quadratically in the step size, $\varepsilon_{\mathrm{step}}$, in the limit $\varepsilon_{\mathrm{step}} \rightarrow 0$ with $N_{\rm step}\varepsilon_{\mathrm{step}}$ held fixed. 
\end{prob} 

\begin{prob}
\label{prob6.2}
Write a function or subroutine that generates initial/final single-nucleon states on the lattice corresponding to a Slater-determinant state with one neutron spin-up and one neutron spin-down, both with zero momentum. \end{prob} 

\begin{prob}
\label{prob6.3}
Write a function or subroutine that generates initial/final single-nucleon
states on the lattice corresponding to a Slater-determinant state with one proton spin-up and one neutron spin-up, both with zero momentum.\end{prob}

\begin{prob}
\label{prob6.4}
Write a function or subroutine that generates initial/final single-nucleon
states on the lattice corresponding to a Slater-determinant state of four
nucleons --- proton spin-up, proton spin-down, neutron spin-up, and neutron
spin-down --- each with zero momentum. \end{prob}

\begin{prob}
\label{prob6.5}
Write a function or subroutine that extends the sample code \ref{label_1}  to repeatedly multiply the auxiliary-field transfer matrix on the left starting from the initial single-nucleon wave functions.  Include the contributions from the auxiliary fields $s$ and $s_I$ as well as the pion
 field $\pi_I$.  
\end{prob}

\begin{prob}
\label{prob6.6}
Write a function or subroutine that extends the sample code \ref{label_2}  to
repeatedly multiply the auxiliary-field transfer matrix on the right starting from the final
single-nucleon wave functions.  Include the contributions from the auxiliary
fields $s$ and $s_I$ as well as the pion
 field $\pi_I$.  
\end{prob}

\begin{prob}
\label{prob6.7}
Use the Slater-determinant states constructed in Probs.~\ref{prob6.2}, \ref{prob6.3}, \ref{prob6.4} as initial and final states.  In each case apply the functions or subroutines written in Prob.~\ref{prob6.5} and Prob.~\ref{prob6.6} with all coupling constants set to zero.  Verify that in each case the initial/final state is the ground state of the non-interacting system with energy equal to zero.
\end{prob}

\begin{prob}
\label{prob6.8}
Use the Slater-determinant states constructed in Probs.~\ref{prob6.2}, \ref{prob6.3},
\ref{prob6.4} as initial and final states.  Using the functions or subroutines written in Prob.~\ref{prob6.5} and Prob.~\ref{prob6.6}, extend the sample code  \ref{label_5} to compute the derivatives of $V(s)$ with respect to $s({\bf n},n_t)$, $s_I({\bf n},n_t)$, and $\pi_I({\bf n},n_t)$.
\end{prob}

\begin{prob}
\label{prob6.9}
Take the code you have written for Prob.~\ref{prob6.8} and complete the remaining steps needed to do hybrid Monte Carlo updates for $s$, $s_I$, and $\pi_I$. Verify that the change in the action produced by the hybrid Monte Carlo update
is scaling quadratically in $\varepsilon_{\mathrm{step}}$ in the limit
$\varepsilon_{\mathrm{step}} \rightarrow 0$ with $N_{\rm step}\varepsilon_{\mathrm{step}}$
held fixed.
\end{prob}   

\begin{prob}
\label{prob6.10}
Take the code you have written for Prob.~\ref{prob6.9} and complete the remaining
steps needed to calculate the energy of the ground state by computing the ratio of the amplitudes $Z(L_t)/Z(L_t-1)$.
\end{prob}

\section{Codes and Benchmarks}

Complete verisons of the codes discussed in this chapter and developed in the exercises can be found online via \href{http://github.com/ManyBodyPhysics/LectureNotesPhysics/tree/master/doc/src/Chapter6-programs}{this link}.  In order to run the codes, one must first copy the corresponding initial/final wavefunctions (waveinit\_1S0.f90, waveinit\_3S1.f90, or waveinit\_He4.f90) into the file waveinit.f90 used by the main program nuclei.f90. The number of nucleons is controlled by the parameter n\_f in input.f90 and must correspond to the number of nucleons in waveinit.f90.  

As an example we show the beginning of the input file input.f90 for a two nucleon state with spatial lattice spacing $a = 1/(100\;{\rm MeV})$, temporal lattice spacing $a_t = 1/(150\;{\rm MeV})$, box size $L = 4a$, and Euclidean time extent $L_t = 6a_t$.  We use an $O(a^{4})$-improved lattice action for the nucleon hopping coefficients, $O(a^{0})$-improved lattice action for the pion hopping coefficients, and $O(a^{0})$-improved lattice action for the pion-nucleon coupling. The coefficient of the $^1S_0$ contact interaction is tuned to the physical $^1S_0$ $n-p$ scattering length and is $-5.615\times 10^{-5}\;{\rm MeV}^{-2}$.  The coefficient of the $^3S_1$ contact interaction is tuned to the deuteron binding energy at infinite volume and is $-6.543\times 10^{-5}\;{\rm MeV}^{-2}$.

\begin{lstlisting}[language=Fortran,caption=Parameter declarations at the beginning of the file input.f90.]
  parameter(n_f = 2)
  parameter(L = 4)
  parameter(Lt = 6)
  parameter(cutoff = 100.D0, temporalcutoff = 150.D0)
  parameter(improveN = 2)
  parameter(improveP = 0)
  parameter(improveD = 0)
  parameter(c1S0_phys = -5.615D-5)
  parameter(c3S1_phys = -6.543D-5)
 
\end{lstlisting}

Using these values for the parameters of the lattice action, we now present some benchmark values which can be used to test the nuclear lattice simulations in the two-nucleon system.  The values presented in these benchmarks are computed using exact calculations of the two-nucleon transfer matrix.  They provide a useful independent check that there are no errors in the Monte Carlo simulations.  In Table~\ref{1S0} we show the energies for the $^1S_0$ spin combination
of two nucleons.  The initial state is one
neutron spin-up and one neutron spin-down, both at zero momentum, for $L=4a$
and various values of
$L_t$.  The energies are extracted by computing the
ratio of amplitudes $Z(L_t)/Z(L_t-1)$ and setting equal to $\exp(-E\alpha_t)$.  

\begin{table}[tbh]
\caption{Benchmark energies for the $^1S_0$ spin combination
of two nucleons.  The initial state is one
neutron spin-up and one neutron spin-down, both at zero momentum, for $L=4a$
and various values of $L_t$.}%
\begin{center}
\begin{tabular}{p{1.5cm}p{1.5cm}}
\hline\noalign{\smallskip}
$L_t$ & ${\rm energy (MeV)}$ \\
\noalign{\smallskip}\svhline\noalign{\smallskip}
$2$ & $-1.0915$  \\
$4$ & $-1.3987$  \\
$6$ & $-1.6209$  \\
$8$ & $-1.7929$  \\ 
$10$ & $-1.9296$  \\
$12$ & $-2.0398$   \\
$14$ & $-2.1291$   \\
$16$ & $-2.2018$   \\
$18$ & $-2.2610$  \\
$20$ & $-2.3094$ \\
\noalign{\smallskip}
\hline\noalign{\smallskip}
\end{tabular}
\end{center}
\label{1S0}%
\end{table}

We show the energies for the $^3S_1$ spin combination
of two nucleons in Table~\ref{3S1}.  The initial state is one
proton spin-up and one neutron spin-up, both at zero momentum, for $L=4a$
and various values of
$L_t$.  The energies are extracted by computing the
ratio of amplitudes $Z(L_t)/Z(L_t-1)$ and setting equal to $\exp(-E\alpha_t)$.

\begin{table}[tbh]
\caption{Benchmark energies for the $^3S_1$ spin combination
of two nucleons.  The initial state is one
proton spin-up and one neutron spin-up, both at zero momentum, for $L=4a$
and various values of
$L_t$.}%
\begin{center}
\begin{tabular}{p{1.5cm}p{1.5cm}}
\hline\noalign{\smallskip}
$L_t$ & ${\rm energy (MeV)}$ \\
\noalign{\smallskip}\svhline\noalign{\smallskip}
$2$ & $-1.4446 $  \\
$4$ & $-2.0400$  \\
$6$ & $-2.4774$  \\
$8$ & $-2.8331$  \\ 
$10$ & $-3.1341$  \\
$12$ & $-3.3925$   \\
$14$ & $-3.6151$   \\
$16$ & $-3.8069 $   \\
$18$ & $-3.9718$  \\
$20$ & $-4.1132$ \\
\noalign{\smallskip}\hline\noalign{\smallskip}
\end{tabular}
\end{center}
\label{3S1}%
\end{table}

\begin{acknowledgement}
The author is grateful for discussions with Amy Nicholson and Morten
Hjorth-Jensen.  He is also greatly indebted to his collaborators Jose
Alarc{\'o}n, Dechuan Du, Serdar Elhatisari, Evgeny Epelbaum, Nico
Klein, Hermann Krebs, Timo L{\"a}hde, Ning Li, Bing-nan Lu, Thomas
Luu, Ulf-G. Mei{\ss}ner, Alexander Rokash, and Gautam Rupak.  Partial
financial support provided by the U.S. Department of Energy
(DE-FG02-03ER41260).  Computational resources were provided by the
J\"{u}lich Supercomputing Centre.
\end{acknowledgement}


\bibliographystyle{spphys}
%
\bibliography{chapter6}


\printindex


\end{document}